\documentclass[useAMS,usenatbib]{mnras}
\usepackage[T1]{fontenc}
\usepackage{ae,aecompl}

\usepackage{mathtext,amssymb,amsmath}
\usepackage{epsfig}
\usepackage{graphics}
\usepackage{url}
\usepackage{adjustbox}
\usepackage{times}

\usepackage{graphicx}	
\usepackage{amsmath}	
\usepackage{amssymb}	
\usepackage{color}
\usepackage[english]{babel}
\usepackage{graphicx}
\usepackage{hyperref}
\usepackage{listings}
\usepackage{lscape}
\usepackage{natbib}
\usepackage{url}
\usepackage{breakurl}
\usepackage{xspace}
\bibpunct{(}{)}{;}{a}{}{,}

\newcommand{\pgs}{PG1654$+$322\xspace}
\newcommand{\pgf}{PG1528$+$025\xspace}

\newcounter{Rco}

\newcommand{\logg}{\mbox{$\log g$}\xspace}

\newcommand{\Teff}{\mbox{$T_\mathrm{eff}$}\xspace}

\newcommand{\ebv}{$E_\mathrm{B-V}$\xspace}

\newcommand{\Lsol}{$L_\odot$}
\newcommand{\Msol}{$M_\odot$}
\newcommand{\Rsol}{$R_\odot$}

\title[Discovery of hot subdwarfs covered with helium-burning ash]
      {Discovery of hot subdwarfs covered with helium-burning ash}

\author[K\@. Werner et al.]{
Klaus Werner,$^{1}$\thanks{E-mail: werner@astro.uni-tuebingen.de}
Nicole Reindl,$^{2}$ 
Stephan Geier$^{2}$
and
Max Pritzkuleit$^{2}$
\\
$^{1}$Institut f\"{u}r Astronomie und Astrophysik, Kepler Center for Astro and
Particle Physics, Universit\"{a}t T\"{u}bingen, Sand 1, 72076 T\"{u}bingen,
Germany\\
$^{2}$Institut f\"ur Physik und Astronomie, Universit\"at Potsdam, Karl-Liebknecht-Stra\ss e 24/25, 14476, Potsdam, Germany
}

\pubyear{2018}

\begin{document}
\label{firstpage}
\pagerange{\pageref{firstpage}--\pageref{lastpage}}
\maketitle
 
\begin{abstract}
Helium rich subdwarf O stars (sdOs) are hot compact stars in a
pre-white dwarf evolutionary state. Most of them have effective
temperatures and surface gravities in the range \Teff =
40\,000--50\,000\,K and \logg = 5.5--6.0. Their atmospheres are helium
dominated. If present at all, C, N,  and O are trace elements. The
abundance patterns are explained in terms of nucleosynthesis during
single star evolution (late helium core flash) or a binary He-core
white dwarf merger. Here we announce the discovery of two hot
hydrogen-deficient sdOs (\pgs and \pgf) that exhibit unusually strong
carbon and oxygen lines. A non-LTE model atmosphere analysis of
spectra obtained with the Large Binocular Telescope and by the LAMOST
survey reveals astonishingly high abundances of C ($\approx 20\%$) und
O ($\approx 20\%$) and that the two stars are located close to the
helium main sequence. Both establish a new spectroscopic class of hot
H-deficient subdwarfs (CO-sdO) and can be identified as the remnants
of a He-core white dwarf that accreted matter of a merging low-mass
CO-core white dwarf. We conclude that the CO-sdOs represent an
alternative evolutionary channel creating PG1159 stars besides the
evolution of single stars that experience a late helium-shell flash.
\end{abstract}




\begin{figure*}
 \centering  \includegraphics[width=0.99\textwidth]{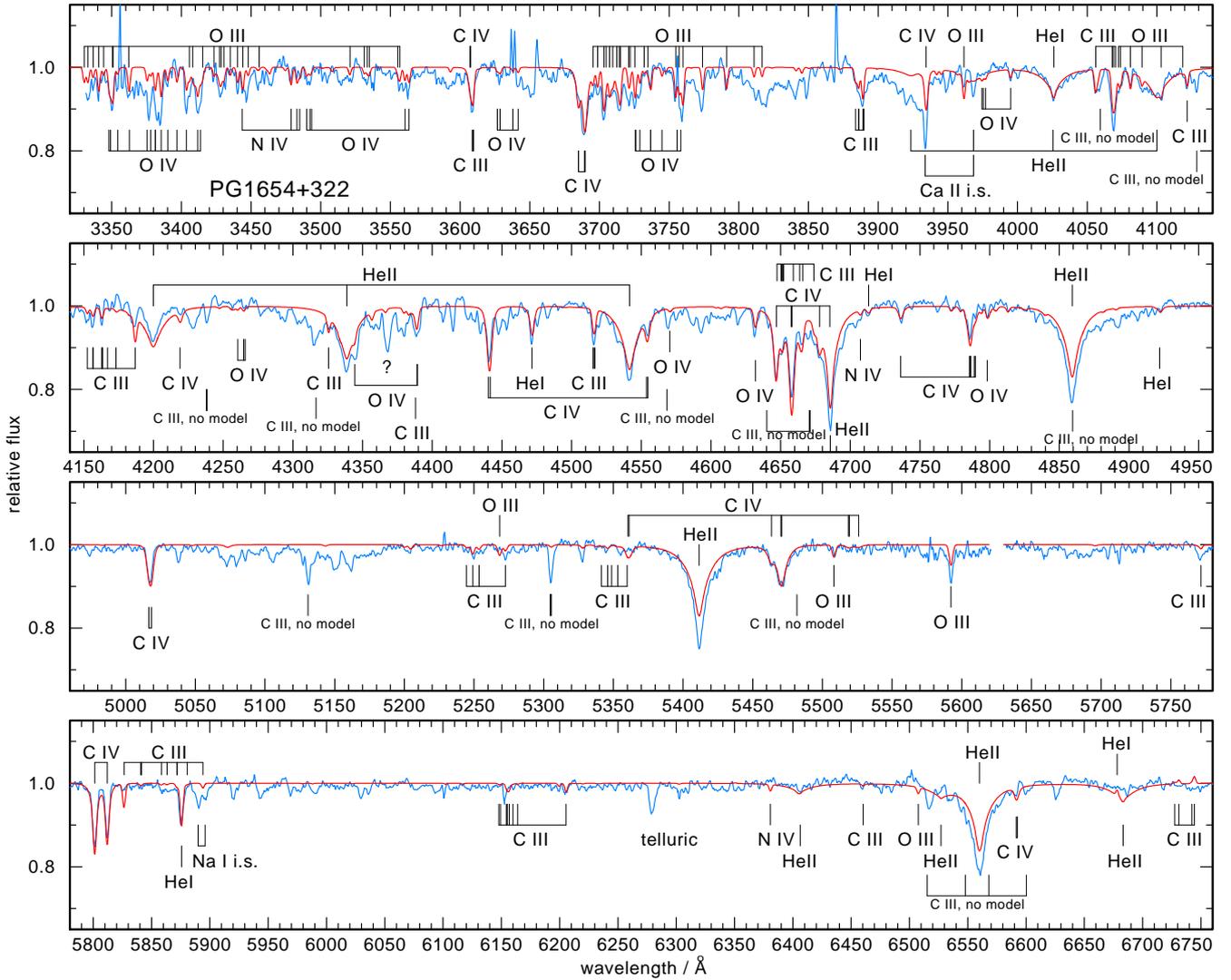}
  \caption{LBT spectrum of \pgs (blue graph). Overplotted is the final
    model (red) with \Teff = 55\,000\,K, \logg = 5.8, and element
    abundances as given in Table\,\ref{tab:resultsall}. Identified
    lines are labeled. Some \ion{C}{iii} lines not included
    in our model atom are visible in the observation and marked with
    ``\ion{C}{iii} no model''. }
\label{fig:pg6}
\end{figure*}

\section{Introduction}
\label{sect:intro}

In the Kiel diagram, helium rich subdwarf O stars (He-sdOs) cluster
around effective temperatures and gravities of \Teff =
40\,000--50\,000\,K and \logg = 5.5--6.0
\citep{2016PASP..128h2001H}. They exhibit little or no hydrogen in
their atmospheres. The carbon abundance is either slightly above the
solar value (up to about C = 0.03 mass fraction) or strongly
depleted. Nitrogen is 3--10 times oversolar for many He-sdOs while it
is strongly depleted in others. Abundance measurements of oxygen are
scarce and reveal values (strongly) below the solar abundance, see
e.g. \citet{2018A&A...620A..36S} and \citet{2021arXiv211113549W}.  The
abundance patterns indicate the dominance of hydrogen-burning ash on
the stellar surface.  As to the origin of He-sdOs, two competing
scenarios are discussed, namely a late He-core flash
\citep[e.g. ][]{2018A&A...614A.136B} or a He-core white dwarf merger
\citep{2012MNRAS.419..452Z}.  The He-sdOs are on or near the helium
main sequence and evolutionary models suggest that they are either
He-core or He-shell burners, respectively. 

In contrast to all He-sdOs studied so far, we present here two
exceptional objects of a new spectral class that we call CO-sdOs. They
caught our attention because they exhibit unusually strong carbon and
oxygen lines in their spectra. We will show that the stars are He rich
with very high C and O abundances, thus, exhibiting a large fraction
of helium-burning ash on their surface. They cannot be explained by
any of the two mentioned evolutionary scenarios.

\pgf was discovered in the Palomar-Green survey
\citep{1986ApJS...61..305G} as a V = 16.4\,mag hot subdwarf and
classified as sdBO. It has been observed twice within the course of
the LAMOST, DR5 \citep{2019ApJ...881....7L,2019ApJ...881..135L}.
These authors classified it as He-sdO and performed an automated
spectral analysis using NLTE models with pure H/He
composition. \cite{2019ApJ...881....7L} found \Teff =
64\,194$\pm$593\,K, \logg = 6.40$\pm$0.08, and a number abundance
ratio $\log {\rm He/H} = 0.55\pm0.14$, while
\cite{2019ApJ...881..135L} arrived at \Teff = 88\,700$\pm$29\,720\,K,
\logg = 5.21$\pm$0.39, and $\log {\rm He/H} = 1.79\pm0.20$.  \pgs was
also discovered in the Palomar-Green survey
\citep{1986ApJS...61..305G} as a V = 15.4\,mag hot subdwarf and
classified as sdOC. No spectroscopic analysis was performed up to now.

\section{Observations and spectral analysis}
\label{sect:obs}

We have observed \pgs with the Large Binocular Telescope (LBT) on July
9, 2021, using the MODS instrument that provides two-channel grating
spectroscopy. The spectra cover the wavelength region
3330--5800\,\AA\ and 5500--10\,000\,\AA\ with a resolving power of
$R\approx1850$ and 2300, respectively. They are shown in
Fig.\,\ref{fig:pg6}. Besides lines of \ion{He}{ii} and weaker
\ion{He}{i} lines, we identify a plethora of carbon lines
(\ion{C}{iii} and \ion{C}{iv}) which are stronger than usual in
He-sdOs. But what is most remarkable is the fact that we see rather
strong oxygen lines (\ion{O}{iii} and \ion{O}{iv}). Usually lines from
this element are very weak if visible at all and can only be detected
in high-resolution spectra. So just by visual inspection it is clear
that these sdOs show unusually high C and O abundances. Nitrogen lines
cannot be detected and there are no hints as to the presence of
hydrogen. \pgf was observed in the course of the LAMOST survey
\citep{2019ApJ...881....7L} on March 21 and April 22, 2017.  The two
spectra do not show any radial velocity shift and the star is
therefore likely not in a close binary system. For the spectral
analysis we use the co-added spectrum. It covers the range
3700--9080\,\AA\ with  $R\approx1800$ at 5500\,\AA\ and looks very
similar to \pgs (Fig.\,\ref{fig:pgf}). To search for variabilities in
the light curves indicative of close binarity we checked the data
archives. Both stars were observed in the course of the ZTF
\citep{2019PASP..131a8003M,2019PASP..131a8002B} 269 and 763 times in
the r-band, respectively. They do not exhibit any variability.

We used the T\"ubingen Model-Atmosphere Package to compute non-LTE,
plane-parallel, line-blanketed atmosphere models in radiative and
hydrostatic equilibrium
\citep{1999JCoAM.109...65W,2003ASPC..288...31W,tmap2012}. We computed
models of the type introduced in detail by
\cite{2014A&A...569A..99W}. They were tailored to investigate the
optical spectra of relatively cool PG1159 stars. In essence, they
consist of the main atmospheric constituents, namely He, C, and O.  N
was included as a trace element in subsequent line-formation
iterations, i.e., the atmospheric structure was kept fixed.

For the analysis of \pgs we computed a small set of models with
different abundances and 5000\,K steps in \Teff\ and 0.3 dex steps in
\logg. The models were calculated in a step-by-step procedure to
improve the spectral fit after each single model
computation. Computing a grid of models by systematically varying all
parameters was prohibitive, because the model atmospheres turned out
to be numerically very unstable in this parameter range.

To constrain the effective temperature we can use the ionisation
balances of He, C, and O. First of all, the relative strengths of
lines from \ion{He}{i} and \ion{He}{ii} are important. In particular,
the neutral helium lines are very sensitive to changes in \Teff. In
addition we can compare lines from \ion{C}{iii} and \ion{C}{iv}, from
which the \ion{C}{iii} lines react strongly on changes in
temperature. Like the \ion{He}{i} lines they become too weak at
too-high temperature. In the case of oxygen, \ion{O}{iii} lines become
weaker with increasing \Teff\ while \ion{O}{iv} lines become stronger
in the temperature region of interest. In the case of \pgs we found a
best fit at \Teff = 55\,000\,K. Since models that are cooler or hotter
by 5000\,K predict the \ion{He}{i} lines much too strong or too weak,
respectively, we estimate our uncertainty for the temperature
determination to $\pm$3000\,K. Constraining the surface gravity is
more difficult because none of the models can fit the lower series
members of the \ion{He}{ii} Pickering lines. In particular, the lines
at 6560, 5412, and 4859\,\AA\ are always too weak in the models. This
is reminiscent of the Balmer line problem
\citep{1993ApJ...407L..85B,1996ApJ...457L..39W} and hence could be a
hint that additional opacity sources that are not included in the
models are important. So we rely on the higher series members. At
too-low gravity the computed line cores become too strong while at
too-high gravity the lines disappear due to pressure broadening. We
find a compromise at \logg = 5.8 and assign a conservatively large
error of $\pm$0.5 dex.

As to the C and O abundances, we started from helium-dominated
atmosphere models with mass fractions of 0.03 for both species and
gradually increased the abundances until a good by-eye fit was
achieved to the spectral lines. We arrived at C = $0.15\pm 0.05$ and O
= $0.23\pm 0.06$. Uncertainties were estimated from models showing too
weak and too strong lines. For nitrogen, only an upper limit could be
found by the absence of \ion{N}{iv} lines in the observed
spectrum. The location of the strongest \ion{N}{iv} lines that would
be detectable above an abundance threshold of N = 0.005 is indicated
in Fig.\,\ref{fig:pg6}. Our best fit model is computed with this value
for the N abundance and some weak lines are visible. Finally we looked
for an upper limit of the hydrogen abundance. Trace hydrogen would
become detectable first by an emission peak of the H$\alpha$ line
close to the core of the respective \ion{He}{ii} line. We found that
at an abundance exceeding H = 0.005 this emission core would produce a
noticeable dent in the \ion{He}{ii} line profile of the model which is
however not observed. We therefore accept this value as an upper
limit.

The best fitting model is displayed in Fig.\,\ref{fig:pg6}. It is
obvious that a number of observed lines are not present in the
model. Some of them are highly excited \ion{C}{iii} lines which are
not included in our model atom. These lines are indicated as
``\ion{C}{iii} no model''. Many other lines remain unidentified, one
of the strongest is at 4368\,\AA. Also, broad and shallow features are
seen, for example at 3800--3850\,\AA, 3910--3930\,\AA, and
5060-5180\,\AA. We can only speculate that these are other
high-excitation lines from \ion{C}{iii} or \ion{O}{iii-iv}. Note that
some of these unidentified lines are also present in \pgf.  We have
looked for lines from other elements (\ion{Ne}{ii}, \ion{Mg}{ii},
\ion{Al}{iii}, \ion{Si}{iii-iv}, \ion{P}{iv-v}, \ion{S}{iv-v},
\ion{Ti}{v}, \ion{Pb}{iv}) discovered by \cite{2018A&A...620A..36S},
\cite{2019A&A...630A.130D}, and \cite{2021A&A...653A.120D} in
high-resolution spectra of He-sdOs, but to no avail.

The analysis of \pgf proceeded in the same way. Its spectrum is
noisier than that of \pgs and it does not reach as far in the blue
wavelength region (Appendix A). As a consequence, the measured element
abundances have larger error estimates. \pgf is cooler than \pgs as
can bee seen by the stronger \ion{He}{i} lines. The shape of the
\ion{He}{ii} line profiles suggest a lower gravity. We determined
\Teff = $50\,000 \pm 3000$\,K, \logg = $5.3 \pm 0.5$, and abundances
as given in Table~\,\ref{tab:resultsall}. Our values for temperature
and gravity are significantly lower than those found by
\cite{2019ApJ...881....7L} mentioned in the Introduction, probably due
to their model atmospheres that are composed of H and He only.

\begin{figure}
 \centering  \includegraphics[width=0.99\columnwidth]{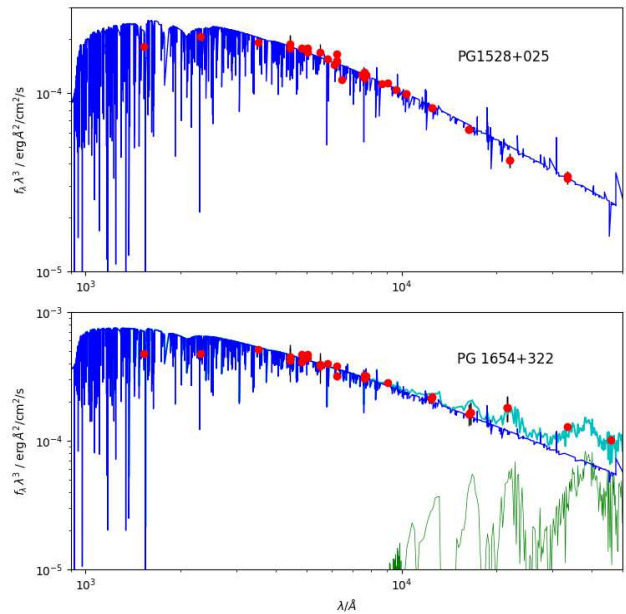}
     \caption{Comparison of our best fitting model fluxes (blue) with
       the observed photometry (red). The IR excess of \pgs is fitted
       with a \Teff = 2000\,K model atmosphere (see text).}
    \label{fig:sed}
\end{figure}

\begin{figure*}
 \centering  \includegraphics[width=0.99\textwidth]{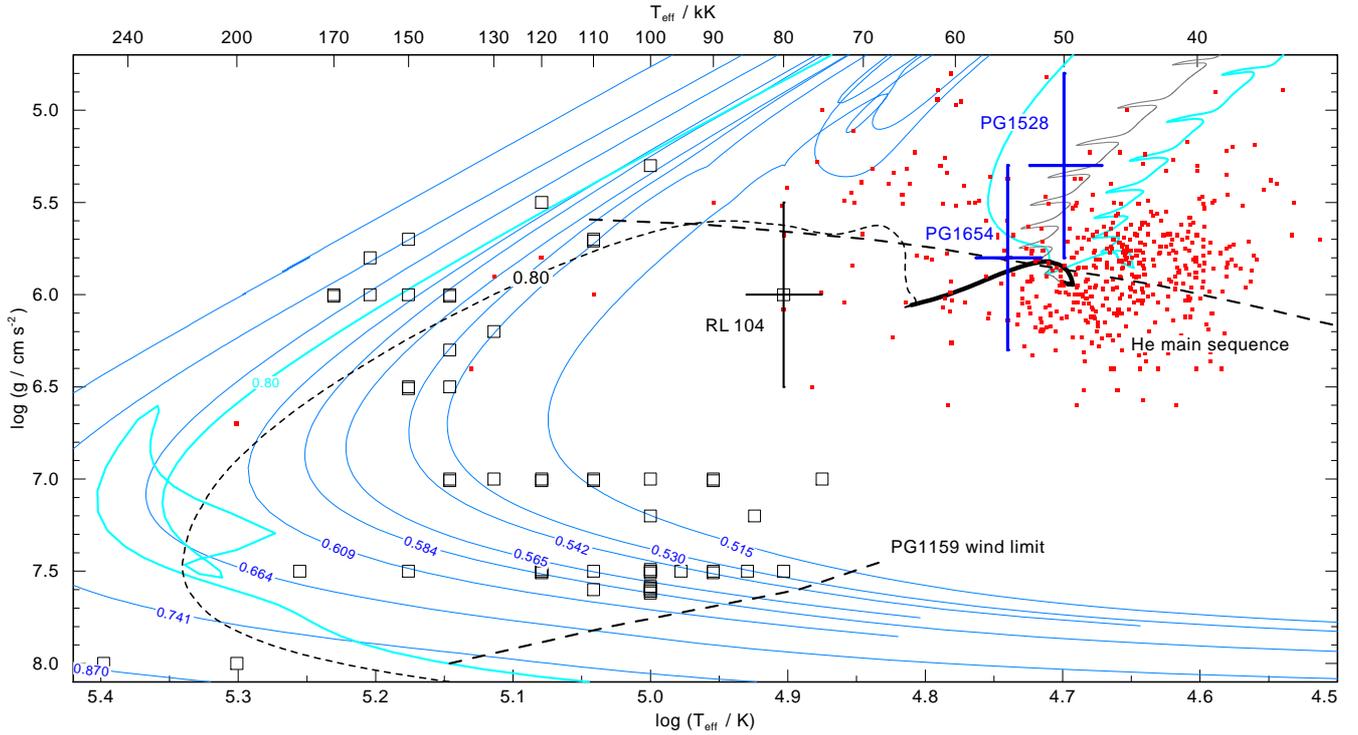}
 \caption{Position of our two CO-sdOs \pgs and \pgf (blue error
   crosses) in the \Teff--\logg\ diagram among He-sdOBs (red symbols)
   and PG1159 stars (black symbols). Blue lines are post-AGB
   evolutionary tracks for late helium-flash stars (marked with
   remnant mass in $M_\odot$). The light blue line is track for a
   merger of two He-core white dwarfs with 0.8\,\Msol\ remnant mass
   ($Z$ = 0.02, He $\approx$ 0.99).  The post-He-core burning phase of
   a 0.8\,\Msol\ remnant of a He+CO white dwarf merger (Z = 0.001, He
   = 0.32, C = 0.20, O = 0.48) is visualized by the thin grey line,
   the He-core burning phase by the thick black line, and the
   post-He-core burning phase by the dashed black line.  The  zero-age
   helium main sequence and the PG1159 wind limit are indicated by
   long-dashed lines.  }
\label{fig:evolution}
\end{figure*}

The model atmospheres for both stars were used to perform a fit to the
observed spectral energy distribution (Fig.\,\ref{fig:sed}).
Using the \cite{Fitzpatrick1999} reddening law our synthetic spectra
were reddened at the values reported by the 2D dust map of
\citet{2011ApJ...737..103S}.  We employed photometry from various
catalogs: GALEX \citep{Bianchi+2017}, Pan-STARRS1
\citep{2016arXiv161205560C}, Landolt B, V \citep{Henden+2015}, Gaia
DR2 and EDR3 \citep{Gaia+2020}, UKIDSS-DR9
\citep{2007MNRAS.379.1599L}, SDSS \citep{2015ApJS..219...12A}, 2MASS
\citep{Cutri2003}, and WISE \citep{Schlafly+2019}. Magnitudes were
converted into fluxes using the VizieR Photometry
viewer (\url{http://vizier.unistra.fr/vizier/sed/}). The
stellar radius follows from the Gaia EDR3 parallax distances from
\citet{2021AJ....161..147B}. The luminosity can then be computed from
$L/L_\odot = (R/R_\odot)^2(T_\mathrm{eff}/T_{\mathrm{eff},\odot})^4$
(Table~\ref{tab:resultsall}).

The visual extinction is evaluated from the standard relation $A_V=3.1
\times$ \ebv so that we obtain a dereddened visual magnitude
$V_0$. The spectroscopic distance $d$ is found by the relation
$$ d {\rm [pc]}= 7.11\times 10^{4} \sqrt{H_\nu\cdot M\cdot 10^{0.4
    V_0-\log g}}\ ,$$
where $H_\nu$ is the Eddington flux of the best-fit atmosphere model
at 5400\,\AA\ (Table~\ref{tab:resultsall}). We assume a mass of
0.8\,\Msol\ (see discussion below). The results are listed in
Table~\ref{tab:resultsall} as well as the distances derived from the
Gaia EDR3 parallaxes. The main error source for the spectroscopic
distance is the uncertainty in the surface gravity, preventing a
meaningful mass determination. Both distance determinations agree
within error limits. 

\pgs shows an infrared excess redward of the SDSS~z-band that can be
reproduced with a NextGen stellar model atmosphere \citep{Allard+2012}
with \Teff = 2000\,K and $R$ = 1.7\,\Rsol.
This radius is more than an order of magnitude higher than what is expected
for an L dwarf ($\approx 0.1$\,\Rsol). For comparison, hot Jupiters and
M dwarfs in a post-common envelope binary with a central star of a
planetary nebula are found to be inflated only by up to $50\%$ (see \citealt{
2020MNRAS.495.2994M,  2020A&A...642A.108J} and references therein). A
close, low-mass companion to \pgs is also doubtful because at such
high primary \Teff, typically emission lines originating from the highly
irradiated side of the cool companion are visible in the optical spectra. A
by-chance alignment of two such peculiar objects also seems unconvincing.
Thus, \pgs may harbor a gaseous disk with a similar temperature.

\begin{table}
\begin{center}
  \caption{ Properties of the analysed hot subdwarfs. Element
    abundances given in mass fractions.  }
\label{tab:resultsall}
\begin{tabular}{rrr}
\hline 
\hline 
                          & \pgs               & \pgf                \\ 
\hline 
\Teff/\,K                 & $55\,000 \pm 3000$ & $50\,000 \pm 3000$ \\
$\log$($g$\,/\,cm\,s$^{-2}$)& $5.8 \pm 0.5$     & $5.3 \pm 0.5$    \\
H                         & $<0.005$           & $<0.005$   \\             
He                        & $0.62\pm 0.11$     & $0.58^{+0.17}_{-0.22}$ \\   
C                         & $0.15\pm 0.05$     & $0.25\pm 0.10$ \\ 
N                         & $<0.005$           & $<0.02$ \\ 
O                         & $0.23\pm 0.06$     & $0.17^{+0.12}_{-0.07}$ \\ 
$L$\,/\,\Lsol             & $418^{+299}_{-167}$  & $1142^{+1960}_{-772}$\\
$R$\,/\,\Rsol             & $0.225^{+0.040}_{-0.030}$ & $0.450^{+0.210}_{-0.160}$ \\
V\,/ mag                  & 15.468             & 16.395               \\
\ebv\,/ mag               & 0.025              & 0.043                \\
$H_\nu$ / erg cm$^{-2}$s$^{-1}$Hz$^{-1}$ & $8.596\times 10^{-4}$ & $7.689\times 10^{-4}$\\
$d$ / pc (spectroscopic)  & $2809^{+2186}_{-1229}$ & $7059^{+5494}_{-3089}$  \\
$d$ / pc (Gaia parallax)  & $3230^{+462}_{-369}$   & $9434^{+4106}_{-3169}$ \\
\hline
\end{tabular} 
\end{center}
\end{table}

\section{Discussion}
\label{sect:discussion}

We have analysed two helium rich (He $\approx$ 0.6) hot subdwarfs
which turn out to be unusually abundant in carbon (C $\approx$ 0.2)
and oxygen (O $\approx$ 0.2; mass fractions). They establish a new
spectral type that we call CO-sdOs in contrast to the usual He-sdOs
which have carbon abundances from strongly subsolar to at most 3\% and
strongly subsolar oxygen abundances
\citep{2009PhDT.......273H,2012MNRAS.427.2180N,2018A&A...620A..36S}.
Fig.\,\ref{fig:evolution} shows the position of our two CO-sdOs in the
Kiel diagram. They are within the region of He-sdOs (red symbols;
\citet{2020A&A...635A.193G,2021MNRAS.501..623J,2021ApJS..256...28L}). This
region is crossed by evolutionary tracks of remnants of a merger of
two He-core white dwarfs.  We show as two examples a
0.5\,\Msol\ remnant (green graph) and a 0.8\,\Msol\ remnant (light
blue graph) from \cite{2012MNRAS.419..452Z}. The He-sdOs and their CNO
abundances are commonly discussed in terms of this merger
scenario. However, the CO-sdOs must have another origin.

The merger of a He-core and a CO-core white dwarf was discussed in the
literature since \cite{1984ApJ...277..355W}. The He white dwarf is
disrupted and accreted onto the CO white dwarf to form a helium
giant. In a companion paper in this volume of MNRAS, Miller Bertolami
et al.\@ suggest that the CO-sdOs discovered by us are the result of
such a merger but in this case the CO white dwarf was the less massive
component being accreted onto the He white dwarf. This can happen only
under very special conditions. First of all the first (stable) mass
transfer episode has to begin close to the tip of the Red Giant
Branch, producing a rather massive He-core white dwarf
($0.4-0.48$\,\Msol). The second requirement is, that the initially
less massive star has to increase its mass to $1.9-2.1$\,\Msol\ during
this first mass transfer event, forming the lowest possible mass sdBs
and CO white dwarfs with masses $0.33-0.4$\,\Msol. In
Fig.\,\ref{fig:evolution} we present the track of a
0.8\,\Msol\ remnant from this scenario. It settles onto the He main
sequence performing He shell flashes before central He burning starts,
causing the small loops. Later on it leaves the He main sequence to
become a hot white dwarf.

The position of our two CO-sdO stars in the Kiel diagram indicates that
they are currently in the He-core burning stage, where CO+He white dwarf
merger products spend most of their pre-white dwarf life ($\approx
22.5$\,Myr). The pre-He-core burning phase lasts for only 0.5\,Myr, yet
stars in that stage are on average ten times more luminous. Thus, for one
pre-He-core burning star 4--5 He-core burning CO-sdO stars can be expected.

After He-core burning, the CO+He-merger evolves through the region of
the PG1159 stars (black symbols in Fig.\,\ref{fig:evolution}). The
abundance patterns of these latter stars (approximately He =
0.30--0.85, C = 0.15--0.60, O = 0.02--0.20) are indeed found to be
very similar to that of the two CO-sdOs. Usually, PG1159 stars are
thought to be post-AGB (pre-) white dwarfs whose H-deficient nature
was caused by a very late thermal pulse (VLTP,
\citet{2006PASP..118..183W}). In Fig.\,\ref{fig:evolution} VLTP
evolutionary tracks from \cite{2006A&A...454..845M} are shown as blue
graphs and the PG1159 wind limit indicating that evolution across this
line transforms PG1159 stars in DA or DO white dwarfs by gravitational
settling of heavy elements \citep{2000A&A...359.1042U}.  However,
while a VLTP object reaches its maximum $T_{\mathrm{eff}}$ within
$\approx 10^4$\,yr, the post-He-core burning and pre-white dwarf stage
(\logg $<$ 7.0) of the CO+He merger lasts for 2.5\,Myr. Considering
that PG1159 stars in the pre-white dwarf stage are about one order of
magnitude more luminous than the CO-sdOs, one could expect one
luminous PG1159 star per CO-sdO in a flux limited star sample.  One
good candidate for this scenario is the recently discovered,
low-luminosity PG1159 star RL104 (black error cross in
Fig.\,\ref{fig:evolution}, \citet{2021arXiv211113549W}), whose
position in the Kiel diagram is not consistent with VLTP tracks. We
conclude that the CO+He white dwarf merger likely represents a
non-negligible evolutionary channel creating PG1159 stars.

One may now speculate if the IR excess of \pgs is a relic from a
circumstellar disk formed by the disrupted CO white dwarf during the
merger event. Such a disk would be made mainly of metals, and it has
been suggested that it could be the birthplace of second generation
planets \citep{2005ApJ...632L..37L}. Therefore, further investigations
on the nature of the IR excess of \pgs are highly desirable.

\section*{Data availability}
Observed and computed spectra are available upon request from the corresponding author.

\section*{Acknowledgements}
We thank Marcelo Miller Bertolami for informing us about his results
on merger evolution calculations explaining the existence of the
CO-sdOs. MP was funded by the Deutsche Forschungsgemeinschaft under
grants GE2506/9-1 and GE2506/12-1. The TMAD tool
(\url{http://astro.uni-tuebingen.de/~TMAD}) used for this paper was
constructed as part of the activities of the German Astrophysical
Virtual Observatory. This research has made use of NASA's Astrophysics
Data System and the SIMBAD database, operated at CDS, Strasbourg,
France. This research has made use of the VizieR catalogue access
tool, CDS, Strasbourg, France. This work has made use of data from the
European Space Agency (ESA) mission Gaia. LAMOST is operated and
managed by the National Astronomical Observatories, Chinese Academy of
Sciences.  Based on observations obtained with the Samuel Oschin
48-inch Telescope at the Palomar Observatory as part of the Zwicky
Transient Facility project, which is supported by the National Science
Foundation under Grant No. AST-1440341 and the participating
institutions of the ZTF collaboration. This paper used data obtained
with the MODS spectrographs built with funding from NSF grant
AST-9987045 and the NSF Telescope System Instrumentation Program
(TSIP), with additional funds from the Ohio Board of Regents and the
Ohio State University Office of Research. The LBT is an international
collaboration among institutions in the United States, Italy and
Germany. LBT Corporation partners are: The University of Arizona on
behalf of the Arizona university system; Istituto Nazionale di
Astrofisica, Italy; LBT Beteiligungsgesellschaft, Germany,
representing the Max-Planck Society, the Astrophysical Institute
Potsdam, and Heidelberg University; The Ohio State University, and The
Research Corporation, on behalf of The University of Notre Dame,
University of Minnesota and University of Virginia.




\bibliographystyle{mnras}
\bibliography{mnras} 



\appendix

\begin{figure*}
 \centering  \includegraphics[width=0.99\textwidth]{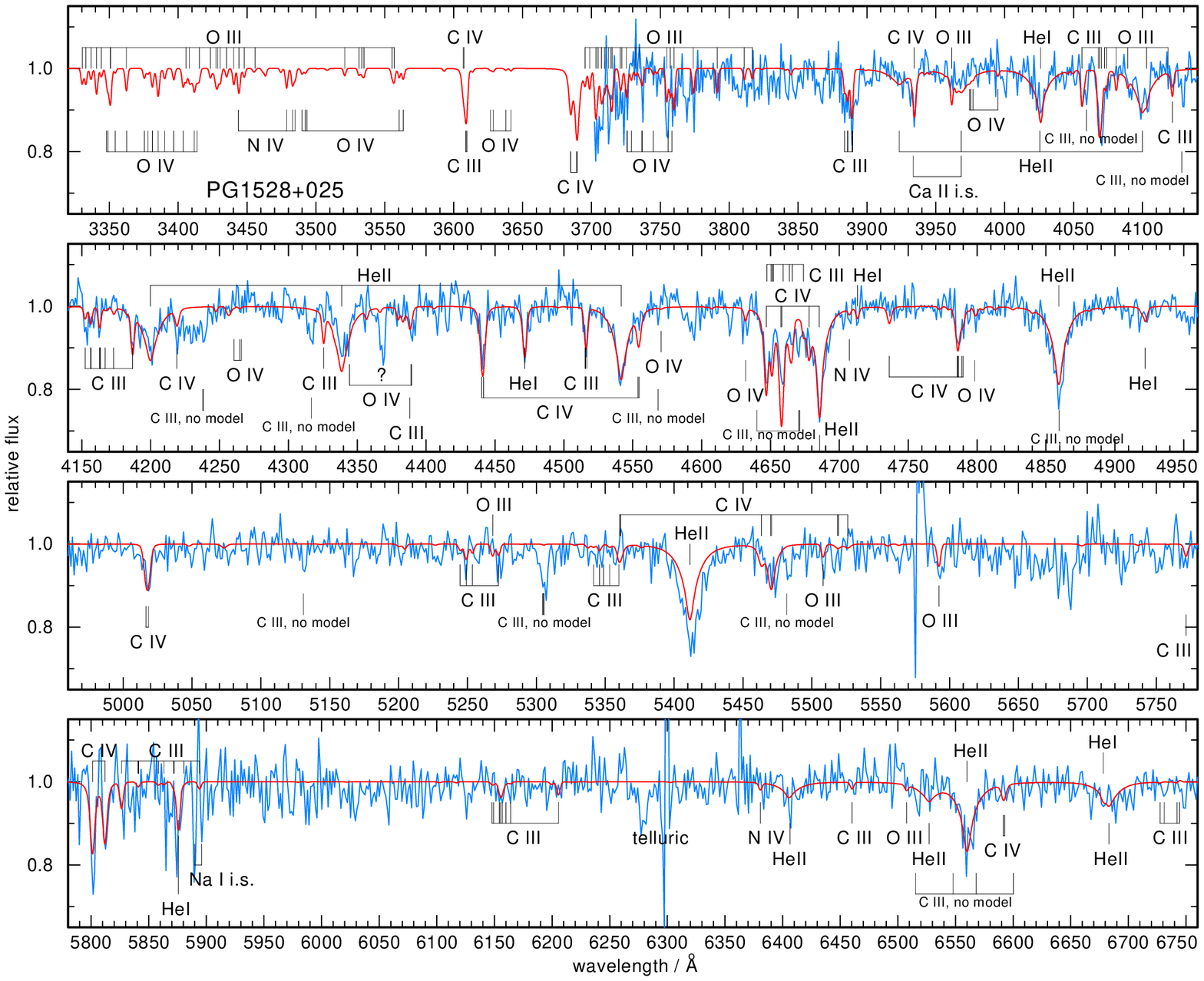}
  \caption{LAMOST spectrum of \pgf (blue
    graph). Overplotted is the final model (red) with \Teff =
    50\,000\,K, \logg = 5.3. }
\label{fig:pgf}
\end{figure*}

\section*{Appendix A}
\label{sect:appendix}

\bsp	
\label{lastpage}
\end{document}